\documentclass[aps,prm,amsmath,amssymb,reprint,
                letterpaper,footinbib,raggedbottom,%
                superscriptaddress,citeautoscript,balancelastpage,%
                floatfix]{revtex4-1}

\usepackage[usenames,dvipsnames]{xcolor}
\usepackage[bookmarks=false,colorlinks]{hyperref}
\hypersetup{
    linkcolor=magenta,        
    citecolor=PineGreen,     
    filecolor=Plum,         
    urlcolor=blue,           
}
\usepackage{graphicx}
\usepackage{mdwlist}
\usepackage[protrusion=true,expansion=true,final]{microtype}



\begin{document}

\title{Inducing Spontaneous Electric Polarizations in Double Perovskite Iodide Superlattices for New Ferroelectric Photovoltaic Materials}
\author{Joshua Young}
	\email{jayoung@binghamton.edu}
	\affiliation{Department\:of\:Physics,\:Binghamton\:University,\:Binghamton,\:NY\:13902,\:USA}
	\affiliation{Department\:of\:Materials\:Science\:and\:Engineering, Drexel\:University, Philadelphia,\:PA 19104, USA}
  \affiliation{Department\:of\:Materials\:Science\:and\:Engineering,\:Northwestern\:University,\:Evanston,\:IL\:60208,\:USA}
\author{James M.\ Rondinelli}
	\email{jrondinelli@northwestern.edu}
  \affiliation{Department\:of\:Materials\:Science\:and\:Engineering,\:Northwestern\:University,\:Evanston,\:IL\:60208,\:USA}
\date{\today}

\begin{abstract}
In this work, we use density functional theory calculations to demonstrate how spontaneous electric polarizations can be induced \textit{via} a hybrid improper ferroelectric mechanism in iodide perovskites, a family well-known to display solar-optimal band gaps, to create new materials for photoferroic applications.
We first assemble three chemically distinct ($A$$A^{\prime}$)($B$$B^{\prime}$)I$_6$ double perovskites using centrosymmetric $AB$I$_3$ perovskite iodides (where $A$ = Cs, Rb, K and $B$ = Sn, Ge) as building units.
In each superlattice, we investigate the effects of three types of $A$- and $B$-site cation ordering schemes and three different $B$I$_6$ octahedral rotation patterns.
Out of these 27 combinations, we find that 15 produce polar space groups and display spontaneous electric polarizations ranging from 0.26 to 23.33 $\mu$C/cm$^2$.
Furthermore, we find that a layered $A$-site/rock salt $B$-site ordering, in the presence of an $a^0a^0c^+$ rotation pattern, produces a chiral ``vortex-like" $A$-site displacement pattern.
We then investigate the effect of epitaxial strain on one of these systems, (CsRb)(SnGe)I$_6$, in layered and rock salt ordered configurations.
In both phases, we find strong competition between the cation ordering schemes as well as an enhancement of the spontaneous polarization magnitude under tensile strain.
Finally, using advanced functionals, we demonstrate that these compounds display low band gaps ranging from 0.2 to 1.3 eV.
These results demonstrate that cation ordering and epitaxial strain are powerful ways to induce and control new functionalities in technologically-useful families of materials. 
\end{abstract}

\maketitle

\section{Introduction}

Solar power is quickly becoming one of the most promising technologies for meeting the world's ever-increasing need for energy.\cite{Bettencourt/Trancik/Kaur:2013}
To this end, new and better materials must be continually developed to keep pace with demand.
Standard semiconducting materials used for photovoltaic technology absorb impinging photons with an energy higher than their band gap, resulting the generation of charge carriers through excitation of electrons into the conduction band (forming corresponding ``holes" in the valence band).
The presence of an asymmetric electric field causes a spontaneous flow of these charge carriers, resulting in a photocurrent that can be harvested as electricity.
Typical photovoltaic devices in production today rely on differences in material composition to transport the charge carriers, such as pn junctions in Si (``homojunctions") or CdTe (``heterojunctions") devices.\cite{Gur_etal:2005,Britt/Ferekides:1993}
However, downsides to this approach include the Shockley-Queisser limit (which states that a single pn junction can not exceed 33.7\% efficiency),\cite{Shockley/Queisser:1961,Xu/Gong/Munday:2015} strong Auger recombination from heavy doping,\cite{Richter_etal:2012} and the need for careful selection of materials to avoid high lattice and band mismatch.
Use of a ferroelectric layer (\textit{i.e.}, a material that displays a spontaneous, switchable electric polarization) for both light absorption/carrier generation and charge separation can help overcome these challenges, as the internal electric field produced by the spontaneous polarization can directly drive charge separation.\cite{Yuan/Xiao/Yang/Huang:2014,Butler/Frost/Walsh:2015}
Moreover, such materials are able to produce photovoltages greater than the band gap, allowing for efficiencies higher than the Schokley-Queisser limit.\cite{Grinberg/Spanier/Rappe_etal:2013}

\begin{figure*}
\centering
\includegraphics[width=1.7\columnwidth,clip]{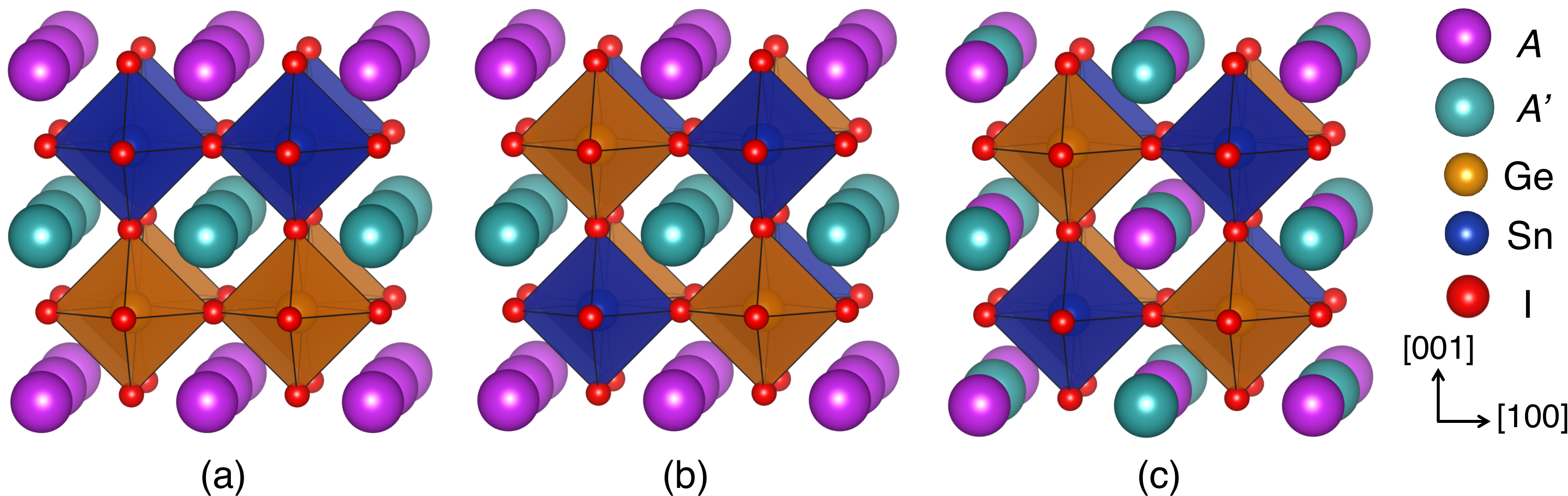}
\caption{The $A$/$A^{\prime}$ and Sn/Ge cations can be ordered along the [001] or [111] axis, resulting in a layered or rock salt scheme, respectively. Within a single ($A$$A^{\prime}$)(SnGe)I$_6$ superlattice, the $A$- and $B$-sites can be ordered in one of three ways: (a) both along the [001] axis ($A_{[001]}/B_{[001]}$), (b) $A$-sites along the [001] axis and $B$-sites along the [111] axis ($A_{[001]}/B_{[111]}$), or (c) both along the [111] axis ($A_{[111]}/B_{[111]}$).
}
\label{fig:orderings}
\end{figure*}

A major challenge in the discovery of new photoferroic materials is finding ferroelectrics with near solar optimal band gaps ($\sim$1.1 eV).\cite{Nechache_etal:2015}
Although oxides and fluorides (in which the vast majority of ferroelectrics are found) have been explored for integration into these types of devices, they typically have wide band gaps of over 3 eV owing to the strong electronegativity of the oxygen and fluorine anions.\cite{Grinberg/Spanier/Rappe_etal:2013}
There has been a surge in attention towards $AB$I$_3$ perovskite iodides over the past few years, as they have small band gaps, high carrier effective masses, and are easily produced through standard synthesis methods.\cite{Snaith:2013,Yin_etal:2015}
Organic-inorganic hybrid materials with organic molecules acting as the $A$-site cation (such as methylammonium lead iodide, where $A$ = CH$_3$CH$_2$NH$_2$ and $B$ = Pb) have reached efficiencies of over 20\%.\cite{Zhou_etal:2014,Yang_etal:2017}
The organic-inorganic hybrids have been predicted to show relatively large spontaneous polarization, but additional verification is needed as the presence of ferroelectricity has been hotly debated (thought to be suppressed by large entropic effects), especially in .\cite{Fan_etal:2015,Alvar_etal:2017,Egger/Rappe_etal:2018,Xiao/Chang_etal:2018}
On the other hand, purely inorganic perovskite iodides such as CsSnI$_3$ have been successfully integrated into photovoltaic devices as well, albeit with lower efficiency.\cite{Kumar_etal:2014,Marshall/Walton/Hatton:2015}
If inversion symmetry can be broken in organic-free iodide perovskites, more robust ferroelectric behavior can be obtained; focusing on these materials also has several other upsides, such as avoiding environmentally unfriendly Pb$^{2+}$ cations.
Several strategies in this vein have recently been utilized for creating low band gap ferroelectrics, including epitaxial strain, charge ordering, and cation ordering across a variety of material families.\cite{Gou/Young/Rondinelli:2017,Wang/Gou/Li:2016,Gou/Charles/Shi/Rondinelli:2017,He/Franchini/Rondinelli:2017,Zhang/Shimada/Kitamura/Wang:2017}

Previous work has shown that ordering of chemically distinct $A$-sites on the atomic scale in perovskites and related phases can lead to a spontaneous polarization arising through a trilinear coupling in the free energy (known as ``hybrid improper" ferroelectricity).\cite{Benedek/Fennie:2011,Rondinelli/Fennie:2012,Young_review:2015}
In this mechanism a combination of purposeful cation ordering and centrosymmetric lattice distortions breaks the inversion symmetry operations preventing the appearance of a spontaneous polarization.
In the $ABX_3$ perovskite family, the most common lattice distortions are cooperative distortions of the anion sublattice, which manifest themselves as ``rotations" of the characteristic $BX_6$ octahedral network.
The octahedra can rotate in-phase (which we characterize by an angle $\Theta_z$, Supplemental Figure 1a) or out-of-phase (characterized by an angle $\Theta_{xy}$, Supplemental Figure 1b) along each of the Cartesian axes.
Glazer notation\cite{Glazer:1972} provides a simple way of describing the different possible combinations of octahedral rotations; a given pattern is indicated by a notation of the form $a^Xb^Xc^X$, where $a$, $b$, and $c$ indicate the respective Cartesian axis, while $X$ describes the type of rotations along that axis ($0$ for no rotations, $+$ for in-phase rotations, and $-$ for out-of-phase rotations).
Finally, if the rotations along two different axes are equal, they are replaced by the same letter. 

So far, the main focus in the field of hybrid improper ferroelectric perovskites has been $A$-site ordering;\cite{Young/Rondinelli_CM:2013,Pilania/Lookman:2014,Young/Rondinelli_PRB:2014,Young/Moon_etal:2017,Young/RondinelliPRB:2016} in this situation, an in-plane polarization arises owing to non-cancelling anti-polar displacements of chemically distinct $A$ and $A^{\prime}$ cations.
While this mechanism is chemistry-independent due to the fact that it is independent of the makeup of the $A$-site and $B$-site cations, further progress has been hindered because $A$-site cation ordering alone can only lift inversion symmetry in the presence of certain rotation patterns, while $B$-site ordering alone can never lift inversion.\cite{Young/Lalkiya/Rondinelli:2016}
In this work, we investigate simultaneous $A$- and $B$-site cation ordering, which can lift inversion symmetry in combination with a much wider array of octahedral rotation patterns.
While some previous work has investigated these types of ordering schemes (such as in double perovskite oxides\cite{Howard/Kennedy/Woodward:2003,Knapp/Woodward:2006,King/Woodward_etal:2007,King/WaymanWoodward:2009,King/Wills/Woodward:2009,Fukushima/Picozzi:2011,Young/Rondinelli_DT:2015}), the focus has primarily been on layered $A$-sites with rock salt ordered $B$-sites.
Here, we consider this ordering scheme, while also investigating double layered and double rock salt ordering of both the $A$- and $B$-site cations, expanding the ways to induce spontaneous polarizations.


Specifically, we utilize density functional theory (DFT) calculations to investigate three chemically distinct ($AA^{\prime}$)(SnGe)I$_6$ double perovskite iodide superlattices ($A$/$A^{\prime}$ = Cs, Rb, and K), with three different types of nanoscale $A$- and $B$-site cation ordering.
Furthermore, we consider three octahedral rotation patterns: $a^0a^0a^0$, $a^0a^0c^+$, and $a^-a^-c^+$, which produce the centrosymmetric cubic $Pm\bar{3}m$, tetragonal $P4/mbm$, and orthorhombic $Pnma$ space groups in bulk $ABX_3$ perovskites, respectively.
We find that a wide variety of non-centrosymmetric phases can be induced in various combinations, which result in many new ferroelectric phases.
We then show how the structure and polarization of different phases of one member of this family [(CsRb)(SnGe)I$_6$] can be tuned using epitaxial strain.

\begingroup
\squeezetable
\begin{table}
\begin{ruledtabular}
\caption{\label{tab:bulk}The space group (s.g.), energy ($\Delta$E), polarization (P), and band gap (E$_g$) of each cation ordered superlattice and the three octahedral rotation patterns investigated. The energy difference between phases is in units of meV normalized to the number of formula units (f.u.), and taken with reference to the lowest energy phase exhibited by each chemistry. The polarization is in units of $\mu$C/cm$^2$, and the axis along which it occurs is given by that axis' unit vector (\^{a}, \^{b}, or \^{c}). The band gap is in eV.}
\begin{tabular}{lcccc}
\multicolumn{5}{c}{(CsRb)(SnGe)I$_6$, $\tau$ = 0.833}  \\
\cline{1-5}
\multicolumn{5}{c}{$A_{[001]}/B_{[001]}$}  \\
\cline{1-5}
Tilt & s. g. & $\Delta$E (meV/f.u.) & P ($\mu$C/cm$^2$) & E$_g$ (eV) \\
\hline\\[-0.6em]
$a^0a^0a^0$ & $P4mm$ & 162.9 & 4.39\:\^{c} & 0.235 \\
$a^0a^0c^+$ & $P4bm$ & 59.96 & 0.80\:\^{c} & 0.468 \\
$a^-a^-c^+$ & $Pc$ & 27.39 & 8.17\:\^{a}, 3.45\:\^{c} & 1.04 \\
\hline
\multicolumn{5}{c}{$A_{[111]}/B_{[111]}$}  \\
\cline{1-5}
Tilt & s.g. & $\Delta$E (meV/f.u.) & P ($\mu$C/cm$^2$) & E$_g$ (eV) \\
\hline\\[-0.6em]
$a^0a^0a^0$ & $P\bar{4}m2$ & 114.2 & -- & 0.734 \\
$a^0a^0c^+$ & $C222$ & 42.87 & -- & 0.642 \\
$a^-a^-c^+$ & $Pc$ & 1.153 & 1.48\:\^{a}, 0.08\:\^{c} & 0.954 \\
\hline
\multicolumn{5}{c}{$A_{[001]}/B_{[111]}$}  \\
\cline{1-5}
Tilt & s. g. & $\Delta$E (meV/f.u.) & P ($\mu$C/cm$^2$) & E$_g$ (eV) \\
\hline\\[-0.6em]
$a^0a^0a^0$ & $P4/nmm$ & 113.1 & -- & 0.461 \\
$a^0a^0c^+$ & $P42_12$ & 54.59 & -- & 0.647 \\
$a^-a^-c^+$ & $P2_1$ & 0 & 16.19\:\^{a} & 0.953 \\
\hline\hline
\multicolumn{5}{c}{(CsK)(SnGe)I$_6$, $\tau$ = 0.815}  \\
\cline{1-5}
\multicolumn{5}{c}{$A_{[001]}/B_{[001]}$}  \\
\cline{1-5}
Tilt & s. g. & $\Delta$E (meV/f.u.) & P ($\mu$C/cm$^2$) & E$_g$ (eV) \\
\hline\\[-0.6em]
$a^0a^0a^0$ & $P4mm$ & 238.3 & 10.49\:\^{c} & 0.282 \\
$a^0a^0c^+$ & $P4bm$ & 120.2 & 0.82\:\^{c} & 0.558 \\
$a^-a^-c^+$ & $Pc$ & 23.69 &11.03\:\^{a}, 3.89\:\^{c} & 1.272 \\
\hline
\multicolumn{5}{c}{$A_{[111]}/B_{[111]}$}  \\
\cline{1-5}
Tilt & s. g. & $\Delta$E (meV/f.u.) & P ($\mu$C/cm$^2$) & E$_g$ (eV) \\
\hline\\[-0.6em]
$a^0a^0a^0$ & $P\bar{4}m2$ & 224.3 & -- & 0.444 \\
$a^0a^0c^+$ & $C222$ & 84.56 & -- & 0.708 \\
$a^-a^-c^+$ & $Pc$ & 5.214 & 2.67\:\^{a}, 0.26\:\^{c} & 1.110 \\
\hline
\multicolumn{5}{c}{$A_{[001]}/B_{[111]}$}  \\
\cline{1-5}
Tilt & s. g. & $\Delta$E (meV/f.u.) & P ($\mu$C/cm$^2$) & E$_g$ (eV) \\
\hline\\[-0.6em]
$a^0a^0a^0$ & $P4/nmm$ & 209.9 & -- & 0.456 \\
$a^0a^0c^+$ & $P42_12$ & 82.79 & -- & 0.702 \\
$a^-a^-c^+$ & $P2_1$ & 0 & 23.33\:\^{a} & 1.104 \\
\hline\hline
\multicolumn{5}{c}{(RbK)(SnGe)I$_6$, $\tau$ = 0.793}  \\
\cline{1-5}
\multicolumn{5}{c}{$A_{[001]}/B_{[001]}$}  \\
\cline{1-5}
Tilt & s. g. & $\Delta$E (meV/f.u.) & P ($\mu$C/cm$^2$) & E$_g$ (eV) \\
\hline\\[-0.6em]
$a^0a^0a^0$ & $P4mm$ & 359.9 & 12.96\:\^{c} & 0.534 \\
$a^0a^0c^+$ & $P4bm$ & 191.9 & 0.91\:\^{c} & 0.570 \\
$a^-a^-c^+$ & $Pc$ & 29.51 & 8.02\:\^{a}, 0.12\:\^{c} & 1.295 \\
\hline
\multicolumn{5}{c}{$A_{[111]}/B_{[111]}$}  \\
\cline{1-5}
Tilt & s. g. & $\Delta$E (meV/f.u.) & P ($\mu$C/cm$^2$) & E$_g$ (eV) \\
\hline\\[-0.6em]
$a^0a^0a^0$ & $P\bar{4}m2$ & 366.1 & -- & 0.420 \\
$a^0a^0c^+$ & $C222$ & 145.2 & -- & 0.684 \\
$a^-a^-c^+$ & $Pc$ & 0.878 & 1.19\:\^{a}, 0.17\:\^{c} & 1.134 \\
\hline
\multicolumn{5}{c}{$A_{[001]}/B_{[111]}$}  \\
\cline{1-5}
Tilt & s. g. & $\Delta$E (meV/f.u.) & P ($\mu$C/cm$^2$) & E$_g$ (eV) \\
\hline\\[-0.6em]
$a^0a^0a^0$ & $P4/nmm$ & 353.9 & -- & 0.425 \\
$a^0a^0c^+$ & $P42_12$ & 136.6 & -- & 0.714 \\
$a^-a^-c^+$ & $P2_1$ & 0 & 16.18\:\^{a} & 1.134 \\
\end{tabular}
\end{ruledtabular}
\end{table}
\endgroup

\section{Computational Methods}

All calculations were performed using density functional theory\cite{Hohenberg/Kohn:1964,Kohn/Sham:1965} as implemented in the Vienna \textit{ab initio} Simulation Package (\texttt{VASP}), using projector augmented wave (PAW) pseudopotentials\cite{Blochl:1994} and the PBEsol functional.\cite{PBEsol:2008}
For the relaxation of the lattice and internal atomic degrees of freedom, we used an 650 eV plane wave cutoff and 7$\times$7$\times$5 Monkhorst-Pack k-point mesh.\cite{Monkhorst/Pack:1976}
The polarization was computed using the Berry phase method and the modern theory of polarization.\cite{King-Smith/Vanderbilt:1993,RevModPhys.66.899}
To simulate the application of epitaxial strain, we fixed the in-plane lattice parameters of each structure to a square strain net and allowed the out-of-plane lattice parameter and internal atomic positions to relax. 
We strained each of the six (CsRb)(SnGe)I$_6$ phases from -10\% to 10\% in increments of 1\%.
%
%
The band gaps were calculated using the HSE06\cite{HSE} hybrid functional with 25\% Hartree-Fock exact exchange included; because of the increased computational cost, we used a reduced 4$\times$4$\times$4 $\Gamma$-centered k-point mesh.

\section{Iodide Superlattices}

We first constructed three chemically distinct ($AA^{\prime}$)($BB^{\prime}$)I$_6$ perovskite superlattices: (CsRb)(SnGe)I$_6$, (CsK)(SnGe)I$_6$, and (RbK)(SnGe)I$_6$.
Of the six $AB$I$_3$ bulk constituents (CsGeI$_3$, CsSnI$_3$, RbGeI$_3$, RbSnI$_3$, KGeI$_3$, and KSnI$_3$), only CsGeI$_3$ and CsSnI$_3$ are known to exhibit the perovskite structure.
CsGeI$_3$ exhibits a rhombohedral $R3m$ structure with a nearly undistorted octahedral network, similar to an undistorted cubic perovskite.\cite{Thiele/Rotter/Schmidt:1987,Krishnamoorthy_etal:2015}
CsSnI$_3$, on the other hand, displays a sequence of three transitions from cubic $Pm\bar{3}m$ ($a^0a^0a^0$) to tetragonal $P4/mbm$ ($a^0a^0c^+$) to orthorhombic $Pnma$ ($a^-a^-c^+$) as a function of decreasing temperature; furthermore, the $Pnma$ perovskite CsSnI$_3$ phase is metastable with an isosymmetric $Pnma$ phase consisting of one-dimensional face-sharing chains of SnI$_6$ octahedra.\cite{Chung/Kanatzidis_etal:2012}
RbGeI$_3$ and RbSnI$_3$ both display this one dimensional chain configuration, as the smaller Rb atoms on the $A$-site are better coordinated in this crystal structure.\cite{Thiele/Rotter/Schmidt:1989,Thiele/Serr:1995}
KSnI$_3$ has been experimentally synthesized, but structural characterization is difficult and a crystal system was unable to be assigned.\cite{Yamada_etal:1990}
Finally, to the best of our knowledge, KGeI$_3$ does not exist and is thermodynamically unstable towards simpler binary compounds.\cite{Krishnamoorthy_etal:2015}
Despite these differences in stability and crystal structure, we chose to constrain our investigations to perovskite superlattices; it seems likely that when the other constituent members are combined in the Cs superlattices, they will be stable in the perovskite form.

\subsection{Bulk phases of superlattices}

We considered three types of cation ordering within each superlattice stoichiometry, resulting in nine distinct compounds: layered $A$-site with layered $B$-sites (which we give the notation $A_{[001]}/B_{[001]}$, Figure \ref{fig:orderings}a), rock salt ordered $A$-sites with rock salt ordered $B$-sites ($A_{[001]}/B_{[111]}$, Figure \ref{fig:orderings}b), and layered $A$-sites with rock salt ordered $B$-sites ($A_{[111]}/B_{[111]}$, Figure \ref{fig:orderings}c).
This notation was selected because layered and rock salt ordering can be thought of as an alternation of chemically distinct cations along the [001] and [111] crystallographic axis, respectively.
Furthermore, these orderings were selected as they are the most amenable to experimental realization of these materials.
The $A_{[001]}/B_{[001]}$ and $A_{[111]}/B_{[111]}$ compounds could be grown \textit{via} layer-by-layer growth of the heterostructure along an [001] or [111] terminated substrate, since layered and rock salt ordering of a particular site can be thought of as alternating the chemically distinct cations along either the [001] or [111] pseudocubic crystallographic axis, respectively.
In terms of spontaneous ordering, the $A_{[001]}/B_{[111]}$ configuration is the most stable in double perovskite oxides; typically, rock salt ordering of the $B$-sites is driven by the need to maximize separation between highly charged cations while layered ordering of the $A$-sites is stabilized by atomic size differences.\cite{King/Woodward:2010}

\begin{figure}
\centering
\includegraphics[width=0.8\columnwidth,clip]{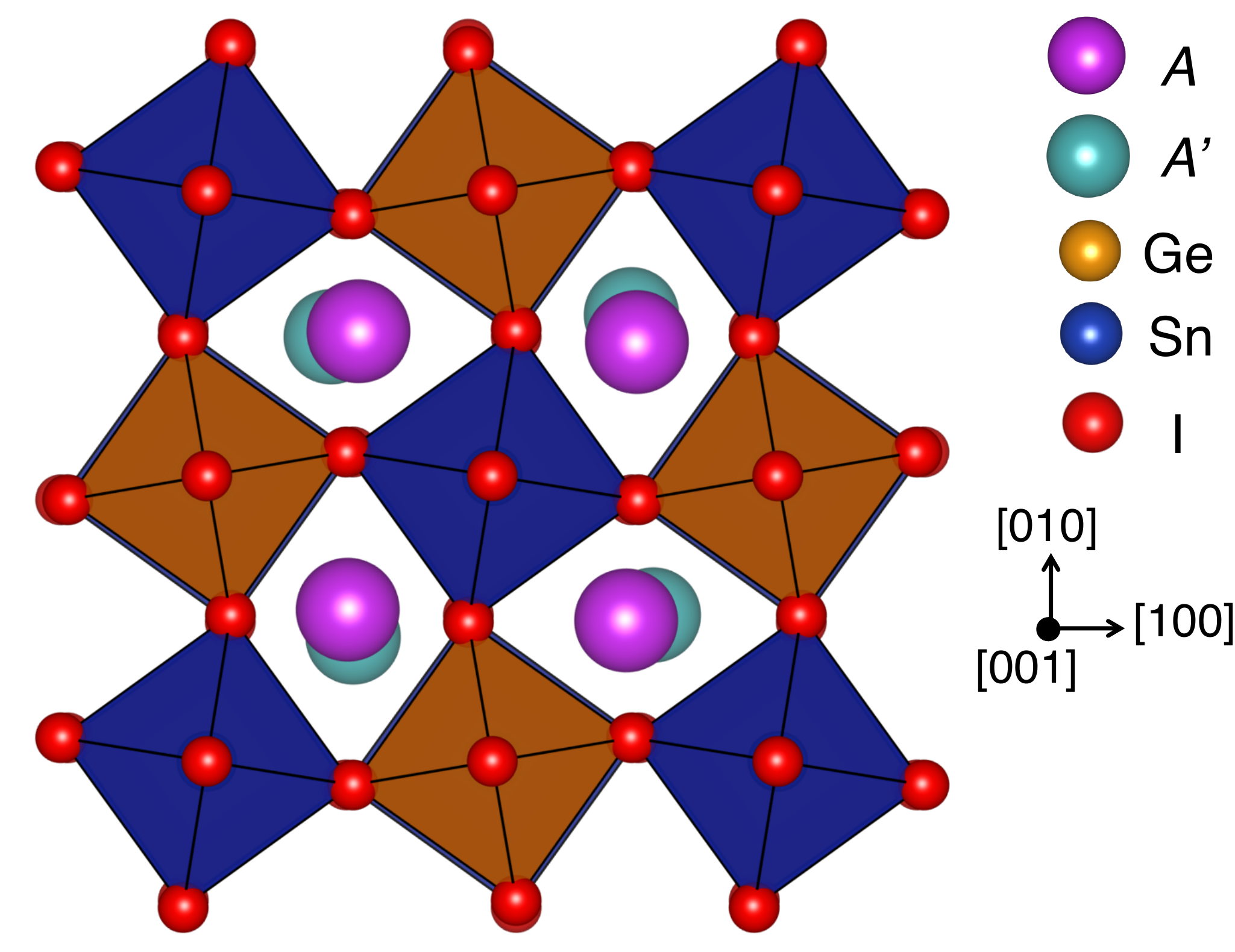}
\caption{A doubling of the unit cell of an iodide superlattice with an $A_{[001]}/B_{[111]}$ ordering scheme and an $a^0a^0c^+$ tilt pattern from 20 to 40 atoms results in ``vortex-like" displacements of the $A$-sites in the $ab$ plane.
}
\label{fig:vortex}
\end{figure}

\subsubsection{Structure and Energetics}

We first fully relaxed the lattice parameters and internal atomic positions of each of the nine superlattices constrained to the $a^0a^0a^0$, $a^0a^0c^+$, and $a^-a^-c^+$ tilt pattern (\textit{i.e.}, those exhibited by bulk CsSnI$_3$), for a total of 27 distinct phases.
The $a$, $b$, and $c$ lattice parameters, as well as the in-phase ($\Theta_z$) and out-of-phase ($\Theta_{xy}$) rotation angles are summarized in Supplemental Table 1.
The energy normalized to the number of formula units (f.u.) of each phase is given by $\Delta$E in Table \ref{tab:bulk}; the lowest energy phase for a given superlattice chemistry is taken to be 0, with the energy of each stoichiometrically equivalent phase taken with reference to it.
In each superlattice, we find the $a^-a^-c^+$ tilt pattern to be lowest in energy, $a^0a^0c^+$ to be the next lowest, and $a^0a^0a^0$ to be the highest; this trend holds regardless of chemical makeup or ordering scheme.
Furthermore, we find that the $A_{[001]}/B_{[111]}$ ordering scheme is the most energetically favorable phase within a given tilt pattern in each of the three chemically distinct superlattices, similar to what is found in oxide double perovskites.\cite{King/Woodward:2010}
The $A_{[111]}/B_{[111]}$ cation ordering scheme is the next most favorable, with the $A_{[001]}/B_{[001]}$ phase being the most energetically unfavorable likely owing to the layered ordering of the $B$-site cations.
However, the energy differences between ordering schemes with a given tilt pattern remain relatively similar across each of the three superlattice compositions.

In the undistorted $a^0a^0a^0$ phase, the octahedra remain unrotated; however, $A$-site displacements along the out-of-plane $c$-axis are allowed, resulting in large spontaneous polarizations which we will discuss in the next section.
When the $a^0a^0c^+$ tilt pattern is enforced, interesting differences appear between the different cation ordering schemes.
In the $A_{[001]}/B_{[001]}$, the SnI$_6$ octahedra are free to rotate independently of the GeI$_6$ octahedra while maintaining corner connectivity; the magnitude of the rotations of these layers are separated in Supplemental Table 1.
The GeI$_6$ octahedra remain nearly unrotated (0.074$^{\circ}$) in $A_{[001]}/B_{[001]}$ (CsRb)(SnGe)I$_6$, while in (CsK)(SnGe)I$_6$ and (RbK)(SnGe)I$_6$ they display larger rotations of 5.64$^{\circ}$ and 11.17$^{\circ}$, respectively; the SnI$_6$ rotation angles remain nearly equivalent across all three chemistries ($\sim$17$^{\circ}$).
When either the $A_{[111]}/B_{[111]}$ or $A_{[001]}/B_{[111]}$ cation ordering scheme is enforced, the GeI$_6$ octahedra are constrained to the in-phase pattern owing to the three dimensional nature of the rock salt patterning scheme of the $B$-sites (in these cases $\Theta_{z,SnO_6}=\Theta_{z,GeO_6}$).
Furthermore, the magnitude of the rotations are reduced when compared to the layered scheme, as the desire of the GeI$_6$ octahedra to remain rigid competes with the SnI$_6$ desire to rotate.

Interestingly, a doubling of the unit cell from 20 to 40 atoms of the $a^0a^0c^+$ phase results in the appearance of a vortex-type pattern of $A$-site displacements. 
This phenomenon occurs across each chemistry, but only in the presence of an $a^0a^0c^+$ tilt pattern and an $A_{[001]}/B_{[111]}$ ordering scheme.
These displacements lift both the $2_1$ screw and the 2-fold rotation axis present in this structure, reducing its symmetry to $P4$; this space group is both polar \textit{and} chiral (whereas $P42_12$ is only chiral), meaning these superlattices display small spontaneous polarizations ($P$, Table \ref{tab:40atom}).
Furthermore, this distortion results in an energy gain ranging from 5 to 10 meV/f.u. ($\Delta$E, Table \ref{tab:40atom}), indicating that this phase is more likely to be the one displayed experimentally.
To quantify the magnitude a given layer ``twists", we define an angle $\Theta_A$ as $\Theta_{A} = \tan^{-1}(d/(a-d))$, where $d$ is the $A$ cations' displacement from its high symmetry position (in \AA) and $a$ is the in-plane lattice parameter.
We find that each $A$-site layer rotates opposite to the $A^{\prime}$ layer above and below it, and that the smaller the $A$-site cation the larger the vortex distortion ($\Theta_A$, Table \ref{tab:40atom}).
This is an interesting effect which deserves further study.

\begingroup
\squeezetable
\begin{table}[t]
\begin{ruledtabular}
\caption{\label{tab:40atom}The energy gain ($\Delta$E) resulting from the $A$-site vortex-type displacements in the 40 atom unit cell of the $A_{[001]}/B_{[111]}$ ordered compounds with an $a^0a^0c^+$ tilt pattern, magnitude of the spontaneous polarization (P) produced, and magnitude of the $A$- and $A^{\prime}$ vortex ($\Theta_A$ and $\Theta_{A^{\prime}}$, respectively). The negative rotation angles indicate that the layers of $A$ and $A^{\prime}$ cations rotate opposite to each other.}
\begin{tabular}{lcccc}
Compound & $\Delta$E (meV/f.u.) & P ($\mu$C/cm$^2$) & $\Theta_A$ ($^{\circ}$) & $\Theta_{A^{\prime}}$ ($^{\circ}$) \\
\hline
(CsRb)(SnGe)I$_6$ & -10.17 & 0.309 & 0.425 & -1.125 \\
(CsK)(SnGe)I$_6$ & -5.194 & 0.339 & 0.609 & -2.600 \\
(RbK)(SnGe)I$_6$ & -4.714 & 0.742 & 1.470 & -2.489 \\
\end{tabular}
\end{ruledtabular}
\end{table}
\endgroup

\begin{figure*}
\centering
\includegraphics[width=1.8\columnwidth,clip]{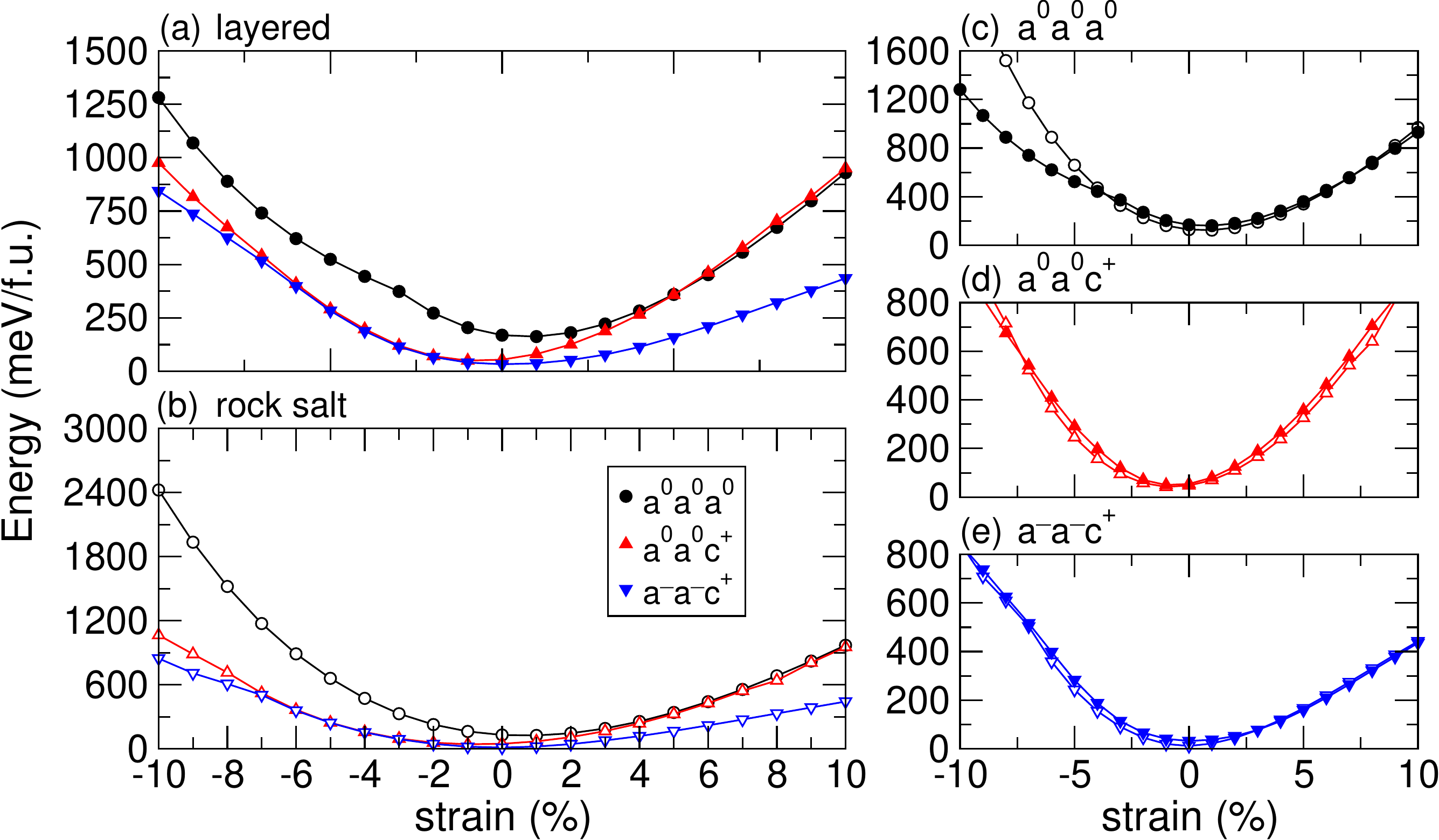}
\caption{Energy as a function of epitaxial strain in (CsRb)(SnGe)I$_6$ with (a) layered and (b) rock salt ordered $A$- and $B$-site cations and an (c) $a^0a^0a^0$, (d) $a^0a^0c^+$, and (e) $a^-a^-c^+$ tilt pattern.
}
\label{fig:tiltenergy}
\end{figure*}

Finally, we investigated the $a^-a^-c^+$ tilt pattern.
As in the $a^0a^0c^+$ case, the SnI$_6$ and GeI$_6$ octahedra are allowed to rotate independently in the presence of an $A_{[001]}/B_{[001]}$ cation ordering scheme in both the in-phase ($\Theta_x$) and out-of-phase ($\Theta_{xy}$) directions.
However, owing to the constraints imposed by the two out-of-phase rotations, the GeI$_6$ octahedra rotate significantly across each chemically distinct $A_{[001]}/B_{[001]}$ compound.
In the $A_{[111]}/B_{[111]}$ or $A_{[001]}/B_{[111]}$ cases the rotation angles are again reduced compared to the layered ordering, with $\Theta_{z,Sn}=\Theta_{z,Ge}$ and $\Theta_{xy,Sn}=\Theta_{xy,Ge}$. 

\subsubsection{Polarization and Electronic Structure}

We next investigated the appearance of ferroelectricity in the polar superlattices; the magnitude of the spontaneous polarization is given as $P$ in Table \ref{tab:bulk}, with the direction given by the unit vector of the corresponding Cartesian axis (\textit{i.e.}, \^{a}, \^{b}, or \^{c}).
In the undistorted $a^0a^0a^0$ phase, only the $A_{[001]}/B_{[001]}$ ordering breaks inversion symmetry, producing a $P4mm$ space group.
In this space group, the $A$-sites sit on the $1b$ Wyckoff position, which allows for a free parameter in the $z$ direction; the $A$ and $A^{\prime}$ sites displace and produce a large spontaneous polarization along that direction.
The chemically distinct superlattices then exhibit different polarizations depending on the size of the $A$ and $A^{\prime}$-site cations. 
In (CsRb)(SnGe)I$_6$ and (CsK)(SnGe)I$_6$ the large Cs atoms remains nearly on their high symmetry position while Rb and K off-center; the smaller K cation displaces more than Rb, resulting in a larger polarization (10.49 vs 4.39 $\mu$C/cm$^2$).
(RbK)(SnGe)I$_6$ displays a larger polarization than either of these compounds (12.96 $\mu$C/cm$^2$), as Rb and K cooperatively displace.
The $A_{[001]}/B_{[111]}$ and $A_{[111]}/B_{[111]}$ orderings in combination with this tilt pattern produce the $P4/nmm$ and $P\bar{4}m2$ space group, respectively, and therefore exhibit no spontaneous polarization.

When the $a^0a^0c^+$ tilt pattern is enforced, all three ordering schemes lift inversion symmetry.
However, as in the undistorted case, only the $A_{[001]}/B_{[001]}$ ordering results in a polar space group ($P4bm$), and is therefore the only one that is allowed to display ferroelectricity.
The magnitude of the spontaneous polarization in this phase is considerably less than the undistorted phase ($<$ 1 $\mu$C/cm$^2$); in this case the $A$ and $A^{\prime}$ cations remain close to their high symmetry positions and displace in an anti-polar fashion, rather than cooperatively.
The $A_{[111]}/B_{[111]}$ and $A_{[001]}/B_{[111]}$ orderings both produce chiral space groups ($C222$ and $P42_12$, respectively).

In the presence of an $a^-a^-c^+$ tilt pattern, all three ordering schemes display a polar space group; both the $A_{[001]}/B_{[001]}$ and $A_{[111]}/B_{[111]}$ ordering scheme exhibit $Pc$, while $A_{[001]}/B_{[111]}$ exhibits $P2_1$.
The six $Pc$ structures each display a small out-of-plane polarization and a larger in-plane polarization.
Consistent with previous work on these types of materials, the magnitude of the polarization is larger in the presence of layered $A$-sites ($A_{[001]}/B_{[001]}$) than rock salt ordered ($A_{[111]}/B_{[111]}$).\cite{Young/Rondinelli_CM:2013}
Within a given ordering scheme, (CsK)(SnGe)I$_6$ displays larger polarizations than (CsRb)(SnGe)I$_6$ or (RbK)(SnGe)I$_6$ (which display similar polarizations) owing to the size difference between the $A$-site species.
The in-plane polarization is a well-known feature of $A$-site cation ordered hybrid improper ferroelectric perovskites, and arises via inequivalent displacements of the chemical species.\cite{Rondinelli/Fennie:2012,Mulder:2012,Mulder/Rondinelli/Fennie:2013}
However, because the presence of $B$-site ordering lifts further symmetry operations than $A$-site ordering alone, small anti-polar out-of-plane $A$-site displacements are now allowed, resulting in the observed out-of-plane polarization.

The $A_{[001]}/B_{[111]}$ scheme produces a $P2_1$ space group, and exhibit spontaneous polarizations arising from a trilinear coupling in the free energy, \textit{i.e.} via a hybrid improper mechanism.\cite{Benedek/Fennie:2011}
%
In agreement with previous studies on perovskites with this type of ordering [cite dalton transactions], we find polarizations along the long in-plane axis.\cite{Fukushima/Picozzi:2011,Young/Rondinelli_DT:2015}
The polarization present in (CsRb)(SnGe)I$_6$ and (RbK)(SnGe)I$_6$ are similar (16.19 and 16.18 $\mu$C/cm$^2$, respectively) because of the similarity in the $A$ and $A^{\prime}$ cation size difference; (CsRb)(SnGe)I$_6$ has a larger polarization of 23.33 $\mu$C/cm$^2$ again owing to the large difference in size between Cs and K.

Finally, we computed the band gap of each superlattice using the advanced HSE06 functional (Table \ref{tab:bulk}).\cite{HSE} 
This results in much better agreement between experimentally measured band gaps and those computed using DFT, which are typically underestimated with standard functionals.\cite{underestimation}
Unsurprisingly, we find that the magnitude of the band gap does not change significantly across the various $A$-site chemistries and ordering schemes, as the character of the band edges consists of Sn, Ge, and I states (Supplemental Figure 2).
However, the band gap is strongly influenced by $B$I$_6$ octahedral rotations, 
In the $a^0a^0a^0$ phases the band gap is small, ranging from 0.235 to 0.534 eV.
The band gap opens slightly in the $a^0a^0c^+$ phases owing to reduced overlap from the octahedral rotations, and range from 0.468 to 0.714 eV.
Finally, the $a^-a^-c^+$ phases display the largest band gaps, ranging from 0.953 to 1.295 eV.
Owing to the fact that band gaps in these phases are close to solar optimal, and that they all display ferroelectric polarizations, the compounds with an $a^-a^-c^+$ tilt pattern would make good ferroelectric photovoltaics.

\subsection{Strained phases}

We next investigated the effect of epitaxial strain on the stabilization of different tilt patterns and ordering schemes in (CsRb)(SnGe)I$_6$.
This compound was selected as we expect it to be the most chemically stable phase due to the aforementioned fact that each of its four bulk constituents exist experimentally.
Furthermore, we consider the $A_{[001]}/B_{[001]}$ and $A_{[111]}/B_{[111]}$ cation ordering configurations, which best correspond to growth conditions as thin films.
We fixed the lattice parameters of the superlattice to be equal to those of a cubic substrate, and strained the $a^0a^0a^0$, $a^0a^0c^+$, and $a^-a^-c^+$ phases of both orderings from 10\% compressive to 10\% tensile in steps of 1\% (see Methods for further details).

\subsubsection{Energetics}

As shown in the previous section, the $a^-a^-c^+$ phase is the lowest in energy in the bulk (CsRb)(SnGe)I$_6$ superlattice regardless of cation ordering, followed by the $a^0a^0c^+$ phase, with the $a^0a^0a^0$ phase the highest.
We first investigate the effect of strain on the relative energetic stability of each tilt pattern within the $A_{[001]}/B_{[001]}$ (Figure \ref{fig:tiltenergy}a) and $A_{[111]}/B_{[111]}$ (Figure \ref{fig:tiltenergy}b) cation ordering scheme.
Under tensile strain, the $a^0a^0a^0$ phase becomes much closer in energy to the $a^0a^0c^+$ phase, and becomes more stable in the $A_{[001]}/B_{[001]}$ scheme at $>$ 6\% strain.
In the case of $A_{[111]}/B_{[111]}$ ordering, although the two tilt patterns become extremely close in energy ($\sim$~20 meV/f.u.), the $a^0a^0c^+$ phase remains lower in energy.
Furthermore, in both ordering schemes, the $a^-a^-c^+$ becomes even more stable than either of the other two tilt patterns under tensile strain (\textit{i.e.}, the energy difference between $a^-a^-c^+$ and the other two tilt patterns increases).
In contrast, under compressive strain, the $a^-a^-c^+$ and $a^0a^0c^+$ begin to become closer in energy; however, the $a^0a^0c^+$ never becomes lower in energy for either cation ordering scheme.

We next investigated the effect of strain on the preferred cation ordering scheme within the $a^0a^0a^0$ (Figure \ref{fig:tiltenergy}c), $a^0a^0c^+$ (Figure \ref{fig:tiltenergy}d), and $a^-a^-c^+$ (Figure \ref{fig:tiltenergy}e) tilt patterns.
In both the $a^0a^0a^0$ and $a^0a^0c^+$ phase, the rock salt-type $A_{[111]}/B_{[111]}$ ordering is slightly lower in energy than the layered $A_{[001]}/B_{[001]}$ ordering from tensile to slightly compressive strain states, in agreement with the bulk case.
Below $-3$\% strain in the $a^0a^0a^0$ tilt pattern and $-7$\% strain in the $a^0a^0c^+$ phase, the $A_{[001]}/B_{[001]}$ ordering becomes more favorable.
Furthermore, the energy of the layered ordering becomes significantly lower than the rock salt ordering in the $a^0a^0a^0$ phase when compared to the $a^0a^0c^+$ phase.
Interestingly, the opposite effect is observed in the $a^-a^-c^+$ phase; here, the $A_{[001]}/B_{[001]}$ ordering is stabilized above 3\% tensile strain, rather than compressive.
However, the energy difference between the two orderings at a given strain state is small (typically on the order of 20 to 30 meV/f.u), indicating they are likely to strongly compete.


\begin{figure}
\centering
\includegraphics[width=1.0\columnwidth,clip]{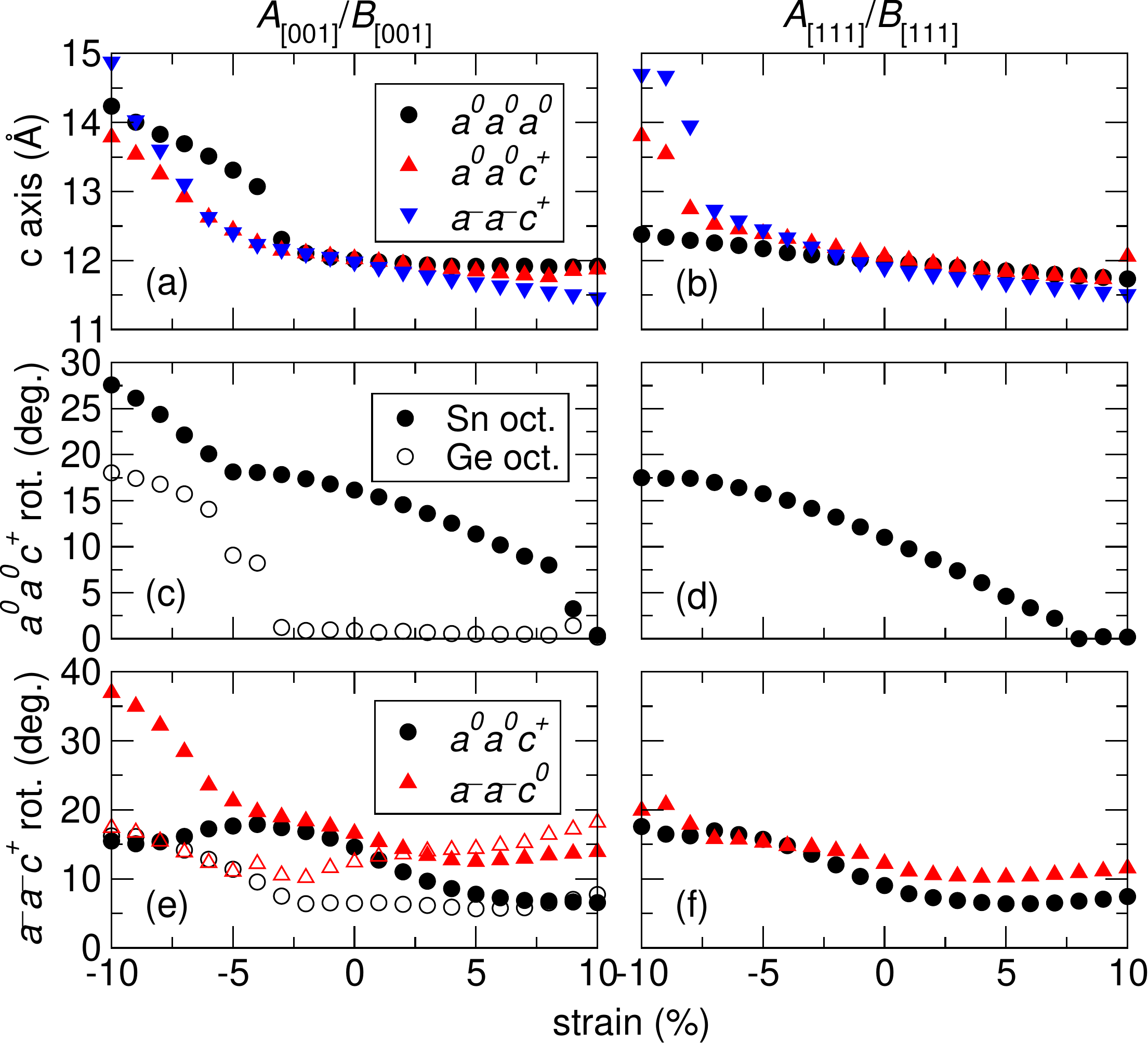}
\caption{Evolution of the lattice parameters of the (a) layered $A_{[001]}/B_{[001]}$ and (b) rock salt $A_{[111]}/B_{[111]}$ ordered structures with the $a^0a^0a^0$ (black circles), $a^0a^0c^+$ (red upward triangles), and $a^-a^-c^+$ tilt pattern (blue downward triangles) as a function of epitaxial strain. Increase in the in-phase rotation angle present in the $a^0a^0c^+$ phase of the (c) $A_{[001]}/B_{[001]}$ and (d) $A_{[111]}/B_{[111]}$ ordered compounds; note that in the layered case, the SnI$_6$ (filled circles) and GeI$_6$ (open circles) octahedra are allowed to rotate independently, while they are forced to be the same in the rock salt case. Evolution of the in-phase (black circles) and out-of-phase (red triangles) rotations in the $a^-a^-c^+$ phase of the (e) $A_{[001]}/B_{[001]}$ and (f) $A_{[111]}/B_{[111]}$ ordered compounds; again, note the separation of SnI$_6$ (filled symbols) and GeI$_6$ (open symbols) in the layered ordering.
}
\label{fig:strainstructure}
\end{figure}

\subsubsection{Structure}

We now seek to explain the difference in energy though an examination of structural changes induced by the strain.
First, we find that the trend in lattice parameters is nearly the same across all three tilt patterns for both the $A_{[001]}/B_{[001]}$ (Figure \ref{fig:strainstructure}a) and $A_{[111]}/B_{[111]}$ ordering scheme (Figure \ref{fig:strainstructure}b).
While the lattice parameter increases with increasing compressive strain as expected, we find sharp increases below $-5$\% in the $A_{[001]}/B_{[001]}$ ordered structures and $-7$\% in the $A_{[111]}/B_{[111]}$ ordered structures; this is likely due to the large size of the iodine atoms, which strongly resist a further decrease in the lattice parameters.
In the $A_{[001]}/B_{[001]}$ case, the SnI$_6$ octahedral layers and GeI$_6$ octahedral layers are allowed to rotate independently without breaking the corner-connectivity of the network  (Figure \ref{fig:strainstructure}c).
Above $-3$\% strain, the GeI$_6$ octahedra remain nearly completely unrotated; this is similar to CsGeI$_3$, which exhibits this same phenomenon.
Furthermore, the magnitude of the SnI$_6$ octahedral rotations increases with increasing compressive strain.
At approximately the same point at which we observe the sharp increase in the out-of-plane lattice parameter, however, we also find sharp increases in the rotation angles as well.
In the $A_{[111]}/B_{[111]}$ ordering, however, the fact that the GeI$_6$ octahedra are surrounded by SnI$_6$ octahedra means that they are forced to rotate.
Note, however, that the magnitude of the rotation angle in the $A_{[111]}/B_{[111]}$ phase at a given strain state is less than that of the SnI$_6$ octahedra in the $A_{[001]}/B_{[001]}$ phase owing to the influence of the GeI$_6$ octahedra being unwilling to rotate.

In the $a^-a^-c^+$ tilt pattern, we can consider the in-phase rotations (which occur about the $c$ axis, $a^0a^0c^+$, black circles) and the out-of-phase rotations (along the $a$ and $b$ axes, $a^-a^-c^0$, red triangles) independently (Figure \ref{fig:strainstructure}e and \ref{fig:strainstructure}f).
Furthermore, as with the previous tilt patterns, the layers of SnI$_6$ (filled points) and GeI$_6$ octahedra (unfilled points) are allowed to rotated independently in the presence of $A_{[001]}/B_{[001]}$ ordering.
As before in this ordering scheme, the magnitude of the in-phase SnI$_6$ rotations increases with increasing compressive strain.
On the other hand, the GeI$_6$ octahedra now rotate slightly rather than remaining unrotated as in the $a^0a^0c^+$ phase; however, they remain relatively unaffected by the strain state until $\sim$ -4\% strain, at which point they begin to increase.
The change in rotation angle is much less pronounced in the $A_{[111]}/B_{[111]}$ phase (Figure \ref{fig:strainstructure}f), with both the in-phase and out-of-phase increasing smoothly as a function of increasing compressive strain.

\subsubsection{Polarization}

\begin{figure}
\centering
\includegraphics[width=0.9\columnwidth,clip]{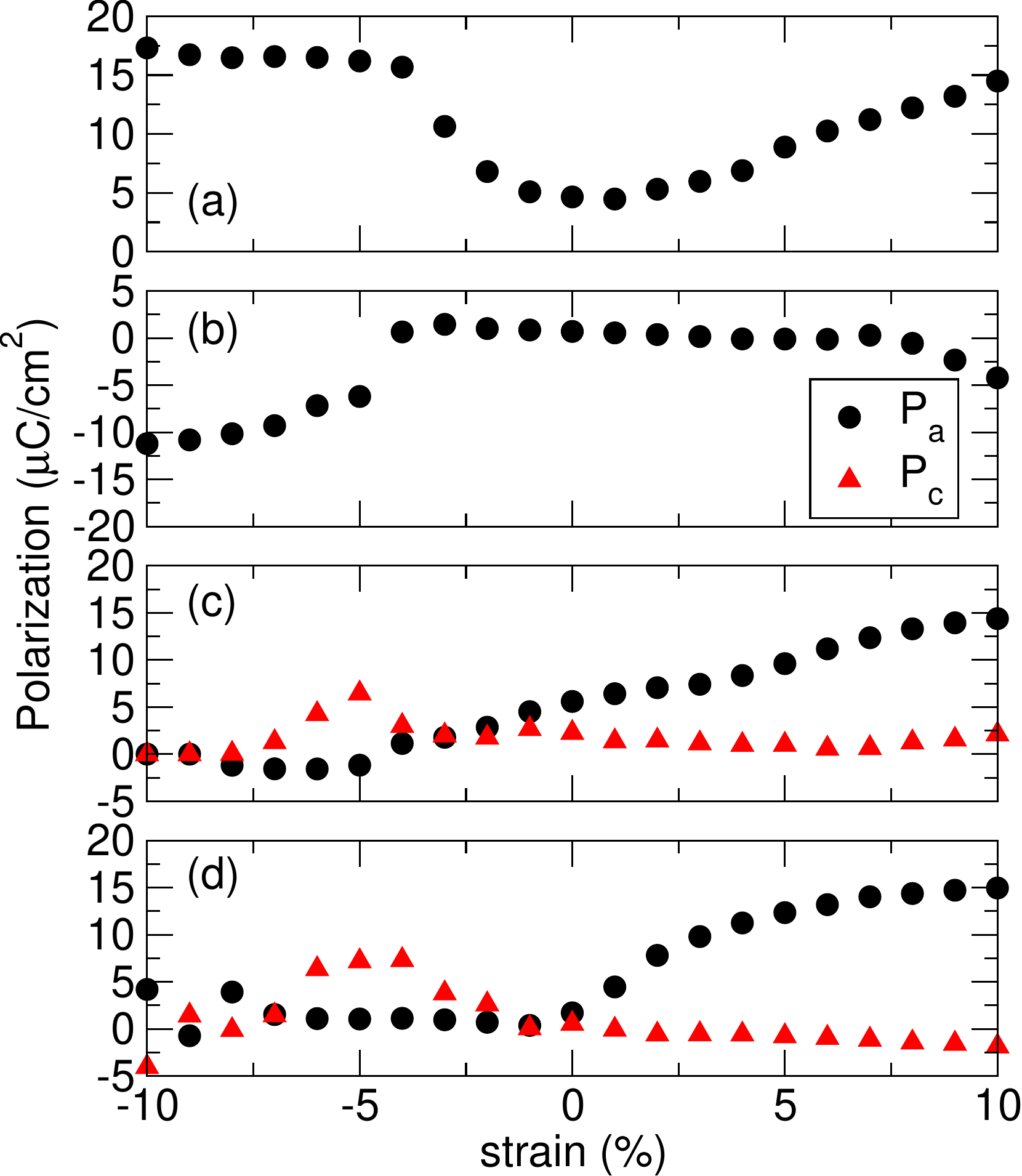}
\caption{The electronic polarization of the $A_{[001]}/B_{[001]}$ cation ordering scheme with the (a) $a^0a^0a^0$, (b) $a^0a^0c^+$, (c) $a^-a^-c^+$ tilt pattern, and (d) the $A_{[111]}/B_{[111]}$ cation ordering scheme with the $a^-a^-c^+$ tilt pattern as a function of strain.
}
\label{fig:polstrain}
\end{figure}

As in the case of bulk (CsRb)(SnGe)I$_6$, the $a^0a^0a^0$ phase of the $A_{[001]}/B_{[001]}$ ordered superlattice under strain displays a relatively large spontaneous polarization resulting from $A$-site displacements along the out-of-plane $c$ axis (Figure \ref{fig:polstrain}a).
The magnitude of this polarization increases under both tensile and compressive strain, and arises via inequivalent displacements of the chemically distinct Rb and Cs $A$-site cations; the polarization primarily results from the significant displacements of the Rb atoms, as the Cs cations remain nearly on their high symmetry positions.
Under tensile strain, $P_c$ linearly increases from 4.66 $\mu$C/cm$^{2}$ at 0\% strain to 14.5 $\mu$C/cm$^{2}$ at 10\%.
However, the situation is quite different under compressive strain, where $P_c$ increases up to -4\% and then saturates at $\sim$17 $\mu$C/cm$^{2}$.

There is then, similar to the bulk case, a large decrease in the polarization when going to the $a^0a^0c^+$ phase (Figure \ref{fig:polstrain}b).
In this case, $P_c$ is 0.71 $\mu$C/cm$^{2}$ at 0\% strain, and does not significantly deviate from this value under tensile strain.
However, there is a sharp jump in the polarization at -5\% compressive strain (-6.19 $\mu$C/cm$^{2}$), again corresponding to the observed spike in the out-of-plane lattice parameter.
The magnitude of the polarization then increases to -11.2 $\mu$C/cm$^{2}$ at -10\% strain.

The orthorhombic phase has two components of the polarization (Figure \ref{fig:polstrain}c), one in-plane ($P_a$) and one out-of-plane ($P_c$).
The in-plane polarization arises via inequivalent displacements of the $A$-site cations and its` magnitude increases under tensile strain, going from 5.60 $\mu$C/cm$^{2}$ to 14.39 $\mu$C/cm$^{2}$ at 0\% and 10\% strain, respectively.
Under compressive strain, however, the magnitude of the polarization decreases and saturates below -4\% strain.
The out-of-plane polarization is slightly smaller (2.27 $\mu$C/cm$^{2}$ at 0\% strain), and remains close to this value over the tensile strain regime.
However, there is some deviation below -4\% compressive strain.

The $A_{[111]}/B_{[111]}$ ordering is only polar in the presence of an $a^-a^-c^+$ tilt pattern, and as such only displays a spontaneous polarization in this phase (Figure \ref{fig:polstrain}d).
As demonstrated in the bulk case, there is an in-plane (P$_a$) and out-of-plane (P$_c$) polarization which both display similar behavior as the $A_{[001]}/B_{[001]}$ ordered phase.
Under tensile strain, we again observe an increase in the in-plane polarization from 4.45 to 14.95 $\mu$C/cm$^{2}$
Unlike the layered case, however, $P_a$ saturates at 0\% strain.
The out-of-plane polarization remains nearly constant as in the layered case, and deviates slightly under compressive strain.

\section{Conclusion}

In this work, we investigated three cation ordered iodide superlattices, (CsRb)(SnGe)I$_6$, (CsK)(SnGe)I$_6$, and (RbK)(SnGe)I$_6$, with three types of cation ordering and three octahedral rotation patterns.
We found that layered $A$- and $B$-site cations ($A_{[001]}/B_{[001]}$) results in a polar space group and spontaneous electric polarization regardless of chemistry and rotation pattern.
In the presence of either double rock salt ordered ($A_{[111]}/B_{[111]}$) or mixed ordered (layered $A$-sites and rock salt ordered $B$-sites, $A_{[001]}/B_{[111]}$) cations, only the $a^-a^-c^+$ tilt pattern produces a ferroelectric polarization.
By investigating the electronic structure of each phase using the advanced HSE06 functional, we found that they display a range of low bands, many of which are close to solar-optimal.
Furthermore, we found that in the presence of in-phase only rotations ($a^0a^0c^+$), the mixed cation ordered phases display a chiral vortex-like $A$-site displacement pattern.
Under epitaxial strain, there is a strong competition between the ($A_{[001]}/B_{[001]}$) and ($A_{[111]}/B_{[111]}$) ordering schemes in (CsRb)(SnGe)I$_6$.
We found that while the $A_{[111]}/B_{[111]}$ ordering scheme is more favorable in bulk and at 0\% strain, the $A_{[001]}/B_{[001]}$ ordering becomes more favorable under large compressive strain in the presence of the $a^0a^0a^0$ and $a^0a^0c^+$ and under large tensile strain in the presence of $a^-a^-c^+$.
We then investigated the evolution of the structural parameters, including out-of-plane lattice parameters and octahedral rotation angles.
Finally, we showed how the spontaneous polarization can be tuned using epitaxial strain, increasing in all phases under tensile strain.
This method of inducing ferroelectricity by ``geometric" (cation ordering and octahedral rotations) rather than chemical effects allows for the judicious selection of material stoichiometry for other properties, such as a solar-optimal band gap.
We hope that these results will provide the impetus for experimental investigation of these materials, especially [001] layer-by-layer growth, which we have shown breaks inversion symmetry in both the presence and absence of octahedral rotations.

\section{Acknowledgements}
J.Y. and J.M.R. acknowledge the National Science Foundation (NSF) for financial support through the Pennsylvania State University MRSEC (DMR-1420620) and CAREER Program (DMR-1454688), respectively.
DFT calculations were performed on the CARBON cluster at the Center for Nanoscale Materials (Argonne National Laboratory, also supported by DOE-BES DE-AC02-06CH11357) under allocation CNM44556 and the Extreme Science and Engineering Discovery Environment (XSEDE), which is supported by NSF Grant No. ACI-1053575.

\bibliography{iodide_strain}
\bibliographystyle{apsrev4-1}

\end{document}